\documentclass[aps,prb,twocolumn,showpacs,showkeys,preprintnumbers,groupedaddress,amsmath,amssymb,]{revtex4-1}

\usepackage{graphicx}
\usepackage{dcolumn}
\usepackage{bm}
\usepackage[dvips]{color}

\begin{document}




\title{Linear temperature behavior of thermopower and strong electron-electron scattering in thick F-doped SnO$_{2}$  films}


\author{Wen-Jing Lang}
\author{Zhi-Qing Li}
\email[Author to whom correspondence should be addressed. Electronic address: ]{zhiqingli@tju.edu.cn}
\affiliation{Tianjin Key Laboratory of Low Dimensional Materials Physics and
Preparing Technology, Department of Physics, Tianjin University, Tianjin 300072,
China}


\date{\today}
\begin{abstract}
Both the semi-classical and quantum transport properties of F-doped SnO$_2$ thick films ($\sim$1\,$\mu$m) were investigated experimentally. It is found that the resistivity caused by the thermal phonons obeys Bloch-Gr\"{u}neisen law from $\sim$90 to 300\,K, while only the diffusive thermopower, which varies linearly with temperature from 300 down to 10\,K, can be observed.The phonon-drag thermopower is completely suppressed due to the long electron-phonon relaxation time in the compound. These observations, together with the temperature independent characteristic of carrier concentration, indicate that the conduction electron in F-doped SnO$_2$ films behaves essentially like a free electron.  At low temperatures, the electron-electron scattering dominates over the electron-phonon scattering  and governs the inelastic scattering process. The theoretical predicated scattering rates for both large- and small-energy-transfer electron-electron scattering processes, which are negligibly weak in  three-dimensional disordered  conventional conductors, are quantitatively tested in this lower carrier concentration and free-electron-like highly degenerate semiconductor.
\end{abstract}

\maketitle

F-doped SnO$_{2}$ (FTO) is one of the typical transparent conducting oxides. Remarkable progress has been made in FTO film deposition techniques recent years.\cite{1,2,3} Currently, both electrical conductivity and optical transparency in visible frequencies of FTO film are comparable to that of Sn-doped In$_{2}$O$_{3}$ (ITO) film.\cite{3,4} Comparing with the most widely used ITO film, FTO film has its own special advantages, such as chemically stable in acidic and basic solutions,\cite{chemically} thermally stable in oxidizing environments at high temperatures,\cite{6,7} and inexpensive (do not include rare elements).  Hence FTO films are widely used in photoelectric and electro-optic devices,  such as solar cells  and flat panel displays.\cite{flat-panel,8,9,10} Although FTO film has been one of the major commercial TCO products, our current understanding of the origins for the combined properties of high electrical conductivity and high optical transparency of FTO film is mainly based on \emph{ab initio} energy bandstructure calculations and optical properties measurements.\cite{4,Frohlich-PRL1978,14,12,13}  Pure SnO$_{2}$ is a wide gap semiconductor with direct bandgap $\sim$$3.6$\,eV and possesses high transmittance in visible light range.\cite{Frohlich-PRL1978,14} The introduction of F ions causes the Fermi level to shift up into the conduction band and the degeneracy of the energy level.\cite{12,13} At the same time, the optical band gap is enlarged comparing with that of pure SnO$_2$ (Burstein-M\"{o}ss effect).\cite{14,12,13}
In addition, the conduction band of FTO is mainly composed of Sn $5s$ state. Hence FTO is a free-electron-like metal or, alternatively a highly degenerate semiconductor in energy bandstructure.\cite{13}  However, the free-electron-like feature of  conduction electrons in FTO has not been tested experimentally. On the other side, the carrier concentrations in  FTO films are often $\sim$$10^{20}$\,cm$^3$,\cite{1,2} which is $\sim$2 to 3 orders of magnitude lower than that in typical metals.\cite{Kittel} The low-carrier-concentration metal characteristic of FTO  may give us opportunities to test the validity of some theoretical predications that is difficult to be achieved in conventional metals. In this letter, We measured the temperature dependence of resistivity and thermopower from 300 K down to liquid helium temperatures, and the results indicate that the transport processes of conduction electrons in FTO films can be approximately treated using free-electron-like model. Then we show that thick FTO film provides a valuable platform to test the three-dimensional (3D) electron-electron (\emph{e}-\emph{e}) scattering theory due to its inherited weak electron-phonon (\emph{e}-ph) coupling nature. It should be noted here that the predications of 3D \emph{e}-\emph{e} scattering theory have not been fully tested though the theory has been proposed for about four decades.\cite{15,16}

FTO (SnF$_{0.06}$O$_{1.94-\delta}$) films prepared by chemical vapor deposition method were provided by Zhuhai Kaivo Optoelectronic Technology Corporation. Two series of films, one is the as-deposited (donated as No.\,1) and the other is the film annealed in O$_{2}$ at 300\,$^\circ$C for 1\,h (denoted as No.\,2), were measured. The thickness of the films ($\sim$$1$\,$\mu$m) was determined by a surface profiler (Dektak, 6\,M). (We intentionally selected the $\sim$$1$\,$\mu$m thick films to make sure they are 3D with respect to \emph{e}-\emph{e} scattering and weak-localization effect.) Crystal structures of the films were measured in a powder x-ray diffractometer (D/max-2500,
Rigaku) with Cu $K_\alpha$ radiation. The results indicated that
the films have tetragonal rutile-type structure, which is the same as that of rutile SnO$_{2}$ (powder diffraction number: 46-1088), and no secondary phase was observed. The resistivity and magnetoresistance (MR) were measured in a physical property measurement system (PPMS-6000, Quantum Design) by a standard four-probe technique. During the MR measurements, the applied field was perpendicular to the films. Hall effect measurements were also performed in the PPMS with the four-point method. The thermopower measurements were carried out with the thermal transport option of the PPMS by a four-probe leads configuration method, in which two calibrated Cernox 1050 thermometers were used to measure the temperature of the hot and cold probes, respectively. The pressure of the sample chamber was less than $5$$\times$$10^{-4}$\,Torr during the measurements.

\begin{figure}[htp]
\begin{center}
\includegraphics[scale=0.75]{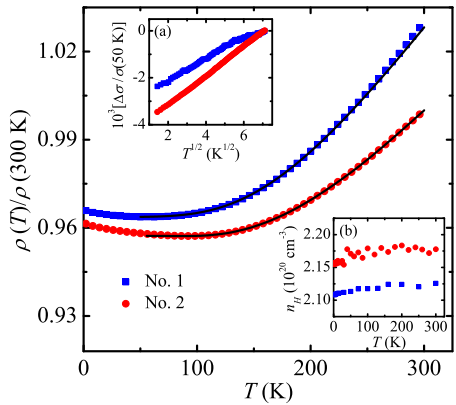}
\caption{Normalized resistivities as a function of temperature for our two FTO films, No.\,1 and No.\,2. The symbols are the experimental data and the solid curves are the least-squares fits to B-G formula. For clarity, the data for film No.\,1 have been shifted by +0.03. Inset (a): variation of normalized conductivity $[\sigma(T)-\sigma(50\,\textup{K})]/\sigma(50\,\textup{K})$ as a function of $T^{1/2}$, and inset (b): carrier concentration obtained from Hall effect measurements vs temperature between 2 and 300\,K.}
\label{FigR-T}
\end{center}
\end{figure}

Figure~\ref{FigR-T} shows the variation in the normalized resistivity $\rho(\emph{T})/\rho(300$\,K) with temperature between 2 and 300\,K for the two FTO films. Upon increasing temperature from 2\,K, the resistivities decrease initially, reach their minimum at $T_{\textup{min}}$ ($T_{\textup{min}}$ is the temperature at which $\rho$ reaches its minimum value, and $T_{\textup{min}}$$\simeq$$50$ and 90\,K for films No.\,1 and No.\,2), then increase with further increasing temperature. The inset (a) of Fig.~\ref{FigR-T} shows the variation of normalized conductivity $\Delta\sigma/\sigma(50\,\textup{K})=[\sigma(T)-\sigma(50\,\textup{K})]/\sigma(50\,\textup{K})$ as a function of $T^{1/2}$ from 2 to $50$\,K. Clearly, $\Delta\sigma$ varies linearly with $T^{1/2}$ at this temperature regime. In 3D disordered metal, the electron-electron interaction (EEI) is strong and leads to $T^{1/2}$ correction to the conductivity.\cite{17,18} The characteristic length for both EEI and \emph{e}-\emph{e} scattering effect is $L_T = \sqrt{\hbar D/(k_B T)}$, where $D$ is the electron diffusion
constant, $\hbar$ is the Planck constant divided by $2\pi$, and $k_B$ is the Boltzmann constant. For films No.\,1 and No.\,2, the electron thermal diffusion lengths at 2\,K are $\approx$31 and $\approx$60\,nm, respectively,  which are far less than the thickness of the FTO films. Hence our FTO films are 3D with regard to EEI effect, and the behavior of $\Delta\sigma (T)\propto T^{1/2}$  at low temperature regime is then attributed to EEI effect. The $\rho(T)$ data at high temperature regime are compared with Matthiessen's rule,\cite{Kittel} $\rho=\rho_0 +\rho (T)$, where $\rho_0$ is the residual resistivity and $\rho (T)$ is the resistivity caused by thermal phonons and is expressed by the Bloch-Gr\"{u}neisen (B-G) formula.\cite{20} The solid lines in Fig.~\ref{FigR-T} are the least-squares fits to the B-G formula. Clearly, the experimental data are overlapped with the theoretical curves, indicating that FTO films possess typical metallic properties in electrical transport properties. The Debye temperatures $\theta_D$ obtained from the fitting processes are 1096 and 1174\,K for films No.\,1 and No.\,2, respectively. The inset (b) of Fig.~\ref{FigR-T} shows temperature dependence of  carrier concentration $n_H$ (we denote the carrier concentration obtained through Hall effect measurement as $n_H$) from $2$ to $300$\,K. The magnitudes of $n_H$ are almost invariable with temperature over the whole measured temperature range (Hall effect measurements also indicate that the main charge carrier is electron in the FTO films). For metals or degenerate semiconductors, activation energy is not required to donate to the charge carriers.\cite{21} Hence the result that $n_H$  is almost independent of $T$ from liquid helium temperatures to 300\,K confirms the metallic transport nature of FTO films.

\begin{figure}[htp]
\begin{center}
\includegraphics[scale=0.75]{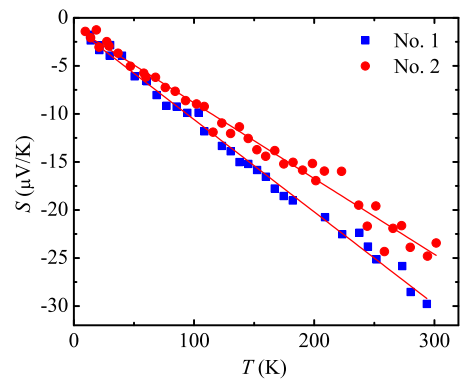}
\caption{Thermopower $S$ as a function of temperature for the two FTO films. The solid straight lines are least-squares fits to Eq.~(\ref{Eq-diffuseS}).}
\label{FigS-T}
\end{center}
\end{figure}

\begin{table*}
\caption{Parameters for the two FTO films, No.\,1 and No.\,2. $\rho$ is resistivity, $n$ is carrier concentration deduced from thermopower measurements. $\emph{A}^S_{ee}$ and $\emph{A}^L_{ee}$ are defined in Eq.~(\ref{Eq-eefit}) and $(A^{S}_{ee})^{th}$ and $(A^{L}_{ee})^{th}$ are predicted by Eq.~(\ref{Eq-ee}). }
\label{Table}
\begin{ruledtabular}
\begin{center}
\begin{tabular}{ccccccc}
Film &   $\rho\,(300$\,K)& $n$& $\emph{A}^S_{ee}$ &$(\emph{A}^S_{ee})^{th}$&$\emph{A}^L_{ee}$&$(\emph{A}^L_{ee})^{th}$\\
       &(m$\Omega$\,cm)  &($10^{20}$/cm$^{3}$)&(K$^{-3/2}$\,s$^{-1}$)&(K$^{-3/2}$\,s$^{-1}$)&(K$^{-2}$\,s$^{-1}$)& (K$^{-2}$\,s$^{-1}$)\\\hline
1&   1.07&     0.95&  $1.38$$\times$$10^{8}$&$9.41$$\times$$10^{7}$&$5.25$$\times$$10^{7}$&$1.03$$\times$$10^{7}$\\
2&   0.28&     1.36&    $4.69$$\times$$10^{7}$&$1.32$$\times$$10^{7}$&$6.16$$\times$$10^{7}$&$9.77$$\times$$10^{6}$\\
\end{tabular}
\end{center}
\end{ruledtabular}
\end{table*}

Figure~\ref{FigS-T} displays the thermoelectric power  $S$ (thermopower) as a function of temperature for the two FTO films from 10 to 300\,K. Clearly, the thermopowers are negative and vary linearly with temperature over the whole measured temperature range. The negative thermopower means the main charge carrier is electron instead of hole, which is identical to the result obtained from Hall effect measurements. In a typical metal, the thermopower generally contains contributions from two separate mechanisms: thermal diffusion of electron and phonon-drag.\cite{22} When a temperature gradient is present in a sample, the electrons from the hotter end will tend to diffuse towards the colder one, then a thermoelectric potential difference between the hotter and colder ends $\Delta V$ will be generated. Then the magnitude of electric field can be written as, $\vec{E}=S\nabla T$. $S\approx -\Delta V/\Delta T$ is the thermopower.  The phonon-drag thermopower originates from the \emph{e}-ph interaction. When the temperature gradient is present, the phonon distribution will no longer be in thermodynamic equilibrium (there will be a heat flow carried by phonons), and this asymmetry characteristic of the temperature will influence the diffusion by the phonon-electron collisions. This is the phonon-drag effect. According to free-electron model, the diffusive thermopower of pure metals at low temperatures ($T \ll \theta_D $) is given by\cite{22}
\begin{equation}{\label{Eq-diffuseS}}
S_d=-\frac{\pi ^2k_B^2T}{3|e|E_F},
\end{equation}
where $e$ is the electron charge and $E_F$ is the Fermi energy. Since the phonon-drag thermopower does not vary linearly with temperature (see further remarks below) and the measuring temperatures are far less than $\theta_D$, we compare our measured $S(T)$ data with Eq.~(\ref{Eq-diffuseS}) and the least-squares-fitted results are plotted as solid lines in Fig.~2.  Using the fitted values of $E_{F}$, we can obtain the carrier concentration $n$ of the samples through $n=(2m^{\ast}E_F)^{3/2}/(3\pi^{2}\hbar^{3})$, an expression also based on the free-electron model. Here $m^{\ast}$ is the effective mass of the carrier and is taken as $0.3m_{e}$ ($m_e$ is the free-electron mass) for FTO.\cite{23} The values of $n$ are listed in Table~\ref{Table}. The values of $n$ are smaller than that obtained by Hall effect measurements, $n_H$. Specifically, $n$ is about one half of $n_H$ for film No.\,1 and three-fourths of $n_H$ for film No.\,2.
In Sb doped SnO$_{2}$, considering that the conduction band is not strictly parabolic and the effective mass increases slightly with increasing occupation of the conduction band (carrier concentration), Egdell \emph{et al}\cite{Egdell-PRB1999,Egdell-JESRP2003} found that the width of the occupied part of the conduction band calculated by using a modified free-electron model can be quantitatively compared with that obtained from ultraviolet photoelectron spectroscopy measurement.\cite{Cox-SSC1982} While assuming that the conduction band is strictly parabolic and  $m^{\ast}$ is fixed, they obtained a width that is far less than the experimental one. For FTO, besides the Sn $5s$ states, both the Sn $5p$  and F $2p$ states also have a little contribution to the conduction band.\cite{12} Hence its energy-momentum dispersion curves in the vicinity of the conduction band minimum are not strictly parabolic shape either.
The underestimate of $n$ in FTO could partly arise from neglecting the variation in $m^{\ast}$ with carrier concentration. On the other hand, Eq.~(\ref{Eq-diffuseS}) and the relation between $E_F$ and $n$ are both derived from the standard free-electron model, hence a slight deviation to the parabolic curve for the conduction band itself could lead to a discrepancy between $n$ and $n_H$.

At low temparatures, the phonon-drag thermopower $S_g$ can be approximately written as,\cite{22}
\begin{equation}{\label{Eq-S}}
S_g \simeq -\frac{C_L}{3n|e|}\frac{\tau_{\rm ph}}{\tau_{\rm ph} + \tau_{{\rm ph}\textup{-}e}},
\end{equation}
where $C_L$ is the heat capacity of the lattice per unit volume, $\tau_{{\rm ph}\textup{-}e}$  is the phonon-electron (ph-\emph{e}) relaxation time, and   $\tau_{\rm ph}$ is the phonon relaxation time for all the other phonon scattering processes. Assuming $\tau_{\rm ph}\ll \tau_{{\rm ph}\text{-}e}$,\cite{24} one can obtain the equation $S_g \simeq S_d \tau_{\rm ph}/(2\tau_{e\textup{-}{\rm ph}})$ by using the energy-balance realtion $\tau_{{\rm ph}\textup{-}e}C_e = \tau_{e\textup{-}{\rm ph}} C_L$,\cite{25} where $\tau_{e\textup{-}{\rm ph}}$ is the \emph{e}-ph relaxation time, $C_e$ is the heat capacity of electrons per unit volume.
Using the relation $\tau_p=3\kappa/(\bar{v}_{s}^2C_L)$ (where $\kappa$ is the thermal conductivity of phonons and $\bar{v}_s$ is the mean velocity of sound) and the experimental data of $C_L$,\cite{26} $\kappa$,\cite{27} and $\bar{v}_s$\cite{28} for SnO$_2$, we obtain the values of
$\tau_{\rm ph}$ are 2.3$\times$$10^{-11}$, 2.2$\times$$10^{-12}$, 1.5$\times$$10^{-12}$, and 9.8$\times$$10^{-13}$ s at 10, 50, 100, and 200\,K, respectively. The theoretical values of $\tau_{e\textup{-}{\rm ph}}$ for film No.\,1 (No.\,2) are 1.0$\times$$10^{-8}$ (3.7$\times$$10^{-8}$), 4$\times$$10^{-10}$ (1.5$\times$$10^{-9}$), 1$\times$$10^{-10}$ (3.7$\times$$10^{-10}$), 2.6$\times$$10^{-11}$\,s (9.4$\times$$10^{-11}$\,s), at the corresponding temperatures, respectively. Here we take $\tau_{e\textup{-}{\rm ph}}$ as $\tau_{e\textup{-}t,\textup{ph}}$, the relaxation time of electron-transverse phonon scattering. (The problem of \emph{e}-ph scattering will be discussed in detail below.) Thus the contribution of $S_g$ to the total thermopower is no more than 2$\%$ of that of $S_d$ at the whole measured temperatures, and can be safely ignored. We note in passing that the temperature dependences of resistivity and carrier concentration indicate that FTO films reveal metallic characteristics in electrical transport properties, while the linear temperature behavior of $S(T)$ as well as the comparability between $n$ and $n_H$ further demonstrate the charge carriers in FTO possess free-electron-like Fermi gas features.

Now we investigate the quantum transport properties of the samples. We note that the drop of the resistivity from 300\,K down to $T_{\rm min}$ is only $\sim$$5\%$ (6.5\% for film No.\,1 and 4.0\% for No.\,2), which indicates the presence of a high level of disorder in the films. This is confirmed by the slight increment of the resistivity below $T_{\rm min}$. The values of disorder parameter $k_F\ell$, deduced from free-electron-like model, are $\approx$6.6 and $\approx$24.1 for fims No.\,1 and No.\,2, respectively, where $k_F$ is the Fermi wave number and $\ell$ is the mean free path of electrons. This indicates that the films fall into the weak-localization region.\cite{25}  In dirty metals and alloys, a lot of investigations have been carried out to detect the electron scattering processes and it has been established that the \emph{e}-ph scattering is the sole dominant inelastic dephasing process in 3D weakly disordered conductors.\cite{25,29}  Recently, Zhang \emph{et al}.\cite{30} found that the small-energy-transfer \emph{e}-\emph{e} scattering can govern the dephasing process in thick ITO films. Their observation demonstrated the validity of the Schmid-Altshuler-Aronov theory of 3D small-energy-transfer \emph{e}-\emph{e} scattering rate in disordered conductors.\cite{15,16} However, the  predication of the theory of 3D large-energy-transfer \emph{e}-\emph{e} scattering rate has not been clearly observed and quantitatively tested up to now. According to Schmid,\cite{15} the total \emph{e}-\emph{e} scattering rate in 3D disordered conductor can be written as,
\begin{equation}
\frac{1}{\tau_{ee}}=\frac{\pi}{8}\frac{(k_B T)^2}{\hbar E_F}+\frac{\sqrt{3}}{2\hbar\sqrt{E_F}}\left( \frac{k_B T}{k_F\ell} \right)^{3/2}.
\label{Eq-ee}
\end{equation}
The first term  on the right hand side of Eq.~(\ref{Eq-ee}) (denoted as $1/\tau_{ee}^L$) represents the contribution of large-energy-transfer \emph{e}-\emph{e} scattering process and would dominate at $k_B T>\hbar/\tau_e$, while the second term (denoted as $1/\tau_{ee}^S$)
stands for the contribution of small-energy-transfer process
and would dominate at $k_B T<\hbar/\tau_e$, where $\tau_e$ is the electron elastic mean free time. Inspection Eq.~(\ref{Eq-ee}) indicates that $1/\tau_{ee}^S\propto (k_F \ell)^{-3/2}$ while $1/\tau_{ee}^L$ is independent of $k_F \ell$. We notice that the $k_F \ell$ values of the ITO films used in
Ref.~\onlinecite{30} range from 1.7 to 3.5, which are much less than that of the FTO films. While the carrier concentrations $n_H$ (or $E_F$) of the FTO films are close to that of the ITO films.  We expect the large-energy-transfer \emph{e}-\emph{e} scattering process, which was not observed in ITO films, would dominate over the small-energy-transfer one at higher temperatures in our FTO films. Then the theoretical predication of total \emph{e}-\emph{e} scattering rate in Eq.~(\ref{Eq-ee}) would be tested.

\begin{figure}
\begin{center}
\includegraphics[scale=0.75]{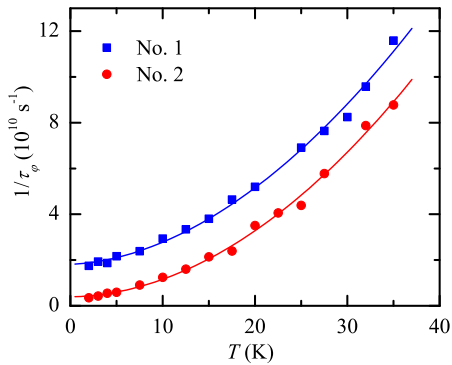}
\caption{variation of $1/\tau_{\varphi}$ with $T$ for the two FTO films. The solid curves are least-squares fits to Eq.~(\ref{Eq-eefit}). For clarity, the data for film No.\,1 are shifted by $+$1$\times$$10^{10}$\,s$^{-1}$.}
\label{Fig-Tau-T}
\end{center}
\end{figure}

To obtain the temperature dependence of electron dephasing rate $1/\tau_{\varphi}$ of the FTO films, we measured the low field MR at different temperatures from 2 to 35\,K. The dephasing rate $1/\tau_{\varphi}(T)$ was then extracted by least-square fitting the MR data to 3D WL theory.\cite{15} The details of the fitting procedure have been discussed previously.\cite{30} Figure~\ref{Fig-Tau-T} shows the variation in $1/\tau_{\varphi}$ with $T$ for the two FTO films, as indicated. We found that only the second term in Eq.~(\ref{Eq-ee}) cannot describe experimental $1/\tau_{\varphi}(T)$ data, i.e., the small-energy-transfer \emph{e}-\emph{e} scattering effect alone cannot explain the electron dephasing process in the FTO films. Here we consider the contributions of both small- and large-energy-transfer \emph{e}-\emph{e} scattering processes and compare our measured $1/\tau_\varphi$ data with the following equation:
\begin{equation}
\frac{1}{\tau_\varphi}=\frac{1}{\tau_\varphi^0}+A_{ee}^S T^{3/2}+A_{ee}^L T^{2},
\label{Eq-eefit}
\end{equation}
where the first, second, and third terms on the right hand side stand for  $T$-independent contribution, small-, and large-energy transfer \emph{e}-\emph{e} scattering rates, respectively. The solid curves in Fig.~\ref{Fig-Tau-T} are the least-squares fits to Eq.~(\ref{Eq-eefit}).  Clearly, the experimental dephasing rate can be well described by Eq.~(\ref{Eq-eefit}). The fitted values of $A_{ee}^S$ and $A_{ee}^L$, together with their theoretical values $(A_{ee}^S)^{th}$ and $(A_{ee}^L)^{th}$ deduced from Eq.~(\ref{Eq-ee}), are listed in Table~{\ref{Table}}. The experimental values of $A_{ee}^S$  ($A_{ee}^L$) are within a factor of $\sim$$4$ ($\sim$$6$) of the theoretical ones. This level of agreement is acceptable.  The value of $A_{ee}^L$ is about one half of that of $A_{ee}^S$ for film No.\,1 and the two values are very close for film No.\,2, which indicates the large-energy-transfer \emph{e}-\emph{e} scattering process has already played important role at liquid helium temperatures in the FTO films.

Besides the large-energy-transfer \emph{e}-\emph{e} scattering, the \emph{e}-ph scattering process also give a $T^{2}$ temperature dependent contribution to the electron dephasing rate. Theoretically, the electron scattering by transverse phonons dominates the \emph{e}-ph relaxation. In the quasi-ballistic limit ($q_T \ell >1$, where $q_T$ is the wavenumber of a thermal phonon), the relaxation rate is expressed as\cite{32,33}  $1/\tau_{e\text{-}t,\text{ph}} = 3\pi^2 k_B^2 \beta_t T^2 / [(p_F u_t)(p_F l)]$, where $\beta_t = (2E_F/3)^2N(E_F)/(2\rho_m u_t^2)$ is the electron--transverse phonon coupling constant,  $p_F$ is the Fermi momentum, $u_t$  is the transverse sound velocity, $\rho_m$ is the mass density, and $N(E_F)$ is the electronic density of states  at the Fermi level. For FTO,
using $u_{t}\approx 3120$\,m/s,\cite{28} one can readily obtain $q_T \ell \approx k_BT\ell/\hbar u_t \approx 0.15T$ and $0.53T$ for films No.\,1 and No.\,2 (the values of $\ell$ are derived using free-electron-like model). Hence our films lie in the quasi-ballistic region above $\sim$$7$\,K. The electronic parameters can also be obtained using free-electron-like model, we take $\rho_m\approx6950$\,kg/m$^3$,\cite{34} the theoretical values of $1/\tau_{e\text{-}t, \rm ph}$ are computed and approximately $9.7\times 10^{5}T^2$ and $2.7\times 10^{5}T^2$\,s$^{-1}$ for films No.\,1 and No.\,2, respectively. Inspection of Table~\ref{Table} indicates the values of $1/\tau_{e\text{-}t, \rm ph}$ are  $\sim$2 order of magnitudes less than the contribution of large-energy-transfer \emph{e}-\emph{e} scattering term. Hence the contribution of \emph{e}-ph relaxation can be safely ignored in the FTO films. In fact, the \emph{e}-ph scattering rate $1/\tau_{e\textup{-}t,\textup{ph}}$ is proportional to the the carrier concentration $n$,\cite{32,33} while Eq.~(\ref{Eq-ee}) predicts $1/\tau_{ee}^S\propto n^{-5/6}$ and $1/\tau_{ee}^L\propto n^{-2/3}$. The carrier concentrations in our FTO films are $\sim$2$\times 10^{20}$\,cm$^3$, which is $\sim$2 to 3 orders of magnitude lower than that in typical metals. Thus the magnitudes of both $1/\tau_{ee}^S$ and $1/\tau_{ee}^L$ are greatly enhanced over the magnitude of \emph{e}-ph scattering rate in the FTO films, which then give us the opportunity to have demonstrated the validity of theories of both the large- and small-energy-transfer \emph{e}-\emph{e} scattering rates in 3D disordered conductors.

In summary, both Boltzmann and quantum-interference transport properties of thick FTO films were investigated experimentally in the the present letter. We found that the experimental $\rho(T)$ data can be well described by the Bloch-Gr\"{u}neisen law from 300\,K down to $T_{\rm min}$, while the carrier concentrations are almost invariable with temperature from 2 to 300\,K. These results, together with linear temperature dependence of thermopowers, demonstrate the conduction electrons in the FTO films possesses free-electron-like characteristics. We also found that both the large- and small-energy-transfer \emph{e}-\emph{e} scattering effect dominate the dephasing process in the measured temperature range (2 to 35\,K) in the FTO films.  Both the linear temperature behavior of $S(T)$ and strong \emph{e}-\emph{e} scattering effect in FTO film are related to the slow \emph{e}-ph relaxation rate, or equally, low carrier concentration characteristic of this highly degenerate semiconductor.

The authors are grateful to Xin-Dian Liu and Pei-Jen Lin for valuable
discussions. This work was supported by the NSF of China through Grant No. 11174216 and Research Fund for the Doctoral Program of Higher Education through Grant No. 20120032110065.


\begin{thebibliography}{00}\label{sec:TeXbooks}
\bibitem{1} T. Maruyama and K. Tabata, J. Appl. Phys. \textbf{68}, 4282 (1990).

\bibitem{2} B. Stjerna, E. Olsson, and C. G. Granqvist, J. Appl. Phys. \textbf{76}, 3797 (1994).

\bibitem{3} A. E. Rakhshani, Y. Makdisi, and H. A. Ramazaniyan, J. Appl. Phys. \textbf{83}, 1049 (1998).

\bibitem{4} E. Shanthi, A. Banerjee, V. Dutta, and K. L. Chopra, J. Appl. Phys. \textbf{53}, 1615 (1982).

\bibitem{chemically} H. Kim, R. C. Y. Auyeung, and A. Piqu\'{e}, Thin Solid Films \textbf{516}, 5052 (2008).

\bibitem{6} B. H. Liao, C. C. Kuo, P. J. Chen, and C. C. Lee, Appl. Opt. \textbf{50}, C106 (2011).

\bibitem{7} J. Ederth, P. Johnsson, G. A. Niklasson, A. Hoel, A. Hult{\aa}ker, P. Heszler, C. G. Granqvist, A. R. van Doorn, M. J. Jongerius, and D. Burgard, Phys. Rev. B \textbf{68}, 155410 (2003).

\bibitem{flat-panel} T. Isono, T. Fukuda, K. Nakagawa, R. Usui, R. Satoh, E. Morinaga, and Y. Mihara, J. Soc. Inf. Disp. \textbf{15}, 161 (2007).

\bibitem{8} H. Kim, G. P. Kushto, R. C. Y. Auyeung, and A. Piqu\'{e}, Appl. Phys. A \textbf{93}, 521 (2008).

\bibitem{9} Z. Hu, J. Zhang, Z. Hao, Q. Hao, X. Geng, and Y. Zhao, Appl. Phys. Lett. \textbf{98}, 123302 (2011).

\bibitem{10} W. H. Baek, M. Choi, T. S. Yoon, H. H. Lee, and Y. S. Kim, Appl. Phys. Lett. \textbf{96}, 133506 (2010).

\bibitem{Frohlich-PRL1978} D. Fr\"{o}hlich, R. Klenkies, and R. Helbig, Phys. Rev. Lett. \textbf{41}, 1750 (1978).

\bibitem{14} G. Sanon, R. Rup, and A. Mansingh, Phys. Rev. B \textbf{44}, 5672 (1991).

\bibitem{12} J. Xu, S. Huang, and Z. Wang, Solid State Commun. \textbf{149}, 527 (2009).

\bibitem{13} J. E. Medvedeva, \emph{Combining Optical Transparency with Electrical Conductivity: Challenges and Prospects}, edited by A. Facchetti and T. J. Marks (Wiley, United Kingdom, 2010).

\bibitem{Kittel} C. Kittel, \emph{Introduction to Solid State Physics} 7th edn (Wiley, New York, 1996).

\bibitem{15} A. Schmid, Z. Phys. \textbf{271}, 251 (1974).

\bibitem{16} B. L. Altshuler and A. G. Aronov, JETP Lett. \textbf{30}, 482 (1979).

\bibitem{17} P. A. Lee and T. V. Ramakrishnan, Rev. Mod. Phys. \textbf{57}, 287 (1985).

\bibitem{18} J. J. Lin and C. Y. Wu, Phys. Rev. B \textbf{48}, 5021 (1993).



\bibitem{20} J. M. Ziman, \emph{Electrons and phonons} (Clarendon Press, Oxford, 1960), p. 364.

\bibitem{21} C. C. Lien, C. Y. Wu, Z. Q. Li, and J. J. Lin, J. Appl. Phys. \textbf{110}, 063706 (2011).

\bibitem{22} D. K. C. MacDonald, \emph{Thermoelectricity: an introduction to the principles} (Wiley, New York, London, 1962).

\bibitem{23} B. Thangaraju, Thin Solid Films \textbf{402}, 71 (2002).

\bibitem{Egdell-PRB1999} R. G. Egdell, J. Rebane, and T. J. Walker, Phys. Rev. B \textbf{59}, 1792 (1999).

\bibitem{Egdell-JESRP2003} R. G. Egdell, T. J. Walker, and G. Beamson, J. Electron Spectrosc. Relat. Phenom. \textbf{128}, 59 (2003).

\bibitem{Cox-SSC1982} P. A. Cox, R. G. Egdell, C. Harding, A. F. Orchard, W. R. Patterson, and P. J. Tavener, Solid State Commun. \textbf{44}, 873 (1982).

\bibitem{24} Since $C_L$ is generally greater than $C_e$ (the heat capacity of electrons per unit volume) above several kelvins and $\tau_{\rm ep}$ is much less than $\tau_{e\textup{-ph}}$ in FTO films, $\tau_{\rm ph}$ is far less than $\tau_{\textup{ph-}e}$.

\bibitem{25} J. J. Lin and J. P. Bird, J. Phys.: Condens. Matter \textbf{14}, R501 (2002).

\bibitem{26} H. Gamsj\"{a}ger, T. Gajda, J. Sangster, S. K. Saxena, and W. Voigt, \emph{Chemical Thermodynamics of Tin} (OECD Publishing, Paris, 2012), p. 420.
\bibitem{27} L. Shi, Q. Hao, C. Yu, N. Mingo, X. Kong, and Z. L. Wang, Appl. Phys. Lett. \textbf{84}, 2638 (2004).

\bibitem{28} A. Di\'{e}guez, A. Romano-Rodr\'{\i}guez, A. Vil\`{a}, and J. R. Morante, J. Appl. Phys. \textbf{90}, 1550 (2001).

\bibitem{29} Y. L. Zhong and J. J. Lin, Phys. Rev. Lett. \textbf{80}, 588 (1998).

\bibitem{30} Y. J. Zhang, Z. Q. Li, and J. J. Lin, Europhys. Lett. \textbf{103}, 47002 (2013).


\bibitem{32} J. Rammer and A. Schmid, Phys. Rev. B \textbf{34}, 1352 (1986).

\bibitem{33} A. Sergeev and V. Mitin, Phys. Rev. B \textbf{61}, 6041 (2000).

\bibitem{34} J. D. Ferguson, K. J. Buechler, A. W. Weimer, and S. M. George, Powder Technol. \textbf{156}, 154 (2005).
\end{thebibliography}
\end{document}